\documentclass[review]{elsarticle}%

\usepackage{lineno,hyperref}
\usepackage{natbib}
\modulolinenumbers[5]
\journal{Journal of ours}
\usepackage{graphicx}
\usepackage{mathrsfs}
\usepackage{amsmath}
\usepackage{amsfonts}
\usepackage[T2A]{fontenc}
\usepackage[cp1251]{inputenc}
\usepackage[russian,english]{babel}
\usepackage{amssymb}%
\usepackage{wasysym}

\newtheorem{theorem}{Theorem}

\newdefinition{remark}{Remark}

\usepackage{setspace}
\usepackage{color}

\newcommand{\blm}{{\boldsymbol{\lambda}}}

\begin {document}
\begin{frontmatter}
\title {On a statistical approach to mate choices in reproduction}
\author {M. Longla}
\address{University of Mississippi\\Department of Mathematics\\ mlongla$@$olemiss.edu}
\author{S. Sivaganesan} \address{ University of Cincinnati\\Department of Mathematical Sciences\\ sivagas$@$ucmail.uc.edu}
\date {}
\begin{abstract}
We provide a probabilistic approach to modeling the movements of subjects  through multiple stages, with "stays" or survival at each stage  for a random length of time,  and ending at a desired final stage. We use  conditional Markov chains with exponential survival times to model the movement of each subject. This  is motivated by a study to learn about of the choices   that different types of  female turkeys make in choosing a male turkey, and in particular, the differences in male choices between groups of females.
In this paper, we propose a model for the subjects' movements toward the final stage, and provide maximum likelihood estimation of the model parameters. We also provide  results relating to certain  questions of   interest, such as the distribution of the number of subjects reaching a stage and the probability that a subject reaches the final stage, and develop  methods for estimating these quantities and  testing statistical hypotheses of interest.

\end{abstract}

\bigskip
\end{frontmatter}

\section{Introduction}
\subsection{Motivation and Problem Setup }
This work is motivated by a study of female turkeys' preferences in mate choice, with the goal of learning about the differences in male choices between groups of females turkeys. In the literature, researchers have mostly considered modeling a function for preferences. Some references to modeling preferences can be found in Kirkpatrick and al (2006), who have proposed a study for mate choice in Tungara frogs. They showed that most of the assumptions are not backed by statistical evidence and suggested that a deeper analysis of the most basic properties of choice rules is necessary.  Dechaume and al. (2013) have shown that female mate choice in convict cichlids is transitive and consistent with a self-referent directional preference. For a better understanding of the topic, one can read Heisler and al. (1987), Navarick and Fantino (1972), Phelps and al. (2006), Bradley and Terry (1952).

{ In this paper, we take a probabilistic approach and propose a  model for data collected in a study on male choice of female turkeys. The study involves a design that monitors movement of female birds through several stages and  the time spent at each stage, until they reach a male bird (the final stage).
  The goal  is to represent the pattern of behavior of female birds by  a parameter that can allow a better representation of the differences between female groups. Assume we have $m$ stages (these are the  stages each female bird may pass through in a transitional horizontal move). There is a finite number of states for each { Stage} $i$ ($i=0,..,m-1$) denoted denoted $s_0, \cdots, s_{m-1}$ (these could be transitional vertical moves of the females through the grid). The individual process moves from a state in Stage $i$ to  any state in Stage $i+1$ with a random waiting time at Stage $i$. The process terminates when a bird reaches the $ m^{th}$ stage or after a time $T=nb$ at the same position, when the individual is removed from the study.}

 The time $T=n\times b$ is the maximal time the process can spend in one given state without move. Moves are checked every $b$ units of time. The underlying assumption is that the process of moves between stages is a conditional Markov chain (conditional on time). This means that for a given set of time stamps, the process of moves works like a Markov chain ignoring the past moves. We will also assume that the distribution of the time conditioned on the set of moves is exponential.

\subsection{Structure of the paper and results}
In Section 2 we provide the likelihood function under the assumptions provided above. The derivations are technical due to the lack of information about the exact time of the moves.

Section 3 presents  maximum likelihood estimators (MLE)  of various quantities of interest and their distributions. For instance, the probabilities to reach the final stage or a specific state of the final stage is provided.
{ The number of records at a given  stage, i.e., the number of times a bird is seen at a given stage is of interest. We derive its probability distribution, and it's mean and variance. A Central Limit Theorem is also provided for the number of entries that reach the final stage. }

A Central Limit Theorem for the maximum likelihood estimator(MLE) of the vector of the probabilities to reach the different states of the final stage is also derived. This  provides a tool for comparison of  the different groups of birds. An alternative estimator of these probabilities based on the multinomial distribution of moves from stage $m-2$ is provided along with the corresponding Central Limit Theorem in formula (\ref{cltKtilde}). For this last case, a limit theorem is proposed for testing inequalities for classification of the states of the final stage within each group. At the end of the section, we provide an alternative way to compare groups through their most probable paths. Here, we propose to use the MLE to estimate the probabilities and compare them.

\section{Distribution and likelihood functions }
 A single observation from this process consists of a set of positions { (sates within stages)} and times { spent at each state}. Additionally, we assume that each of the exponential distributions representing the probability distribution of time spent at a state within a stage depends only on the states and { and not on?} the stages. This means, for example, that the conditional probability distribution function of the time till move to state $j$ of stage $i$ from any state of stage $i-1$ for a path from group $k$ is $ \displaystyle f(t)=\lambda^{k}_{i} e^{-\lambda^{k}_{i}t}$.
%  Its conditional cumulative distribution function $F(t)=1-e^{-\lambda^{k}_{i}t}$).
   We are assuming that the average waiting time does not depend on the state the path is moving to { and the stage it is moving from}, because the impact of the state is incorporated into the transition probabilities. We also assume that the sum of the transition probabilities of the moves to the next stage is equal to $1$.
Our goal is to study the differences between the groups based on their patterns of moves.

 We assume that there are $K$ groups of birds.   Let $X_{\ell,k}$ represent the vector of states visited by the $\ell^{th}$ path from group $k$, where $\ell=1,..,N_k$ and $k=1,...,K$, { where $K$ is the number of groups and $N_k$ is the number of paths from group $k$}. We use
% $(n)_{ik}=(n^{k}_{u,i,1}, \cdots, n^{k}_{u,i,s_{i}})$
 {  $(n)_{ik}=(n^{k}_{u,i,j})$  be the matrix (of size $s_{i-1}\times s_i$) of the numbers of moves  from stage $i-1$ for observations from group $k$},  and $p_{u,i,j,k}$ for the probability of  move from state u of stage $i-1$ to state $j$ of stage $i$ for $i>0$ .
%The index $k$ everywhere in this paper represents the group from which the observation is taken.  There are $K$ such groups.
 Let { $P_{i,k}=( p_{u,i,j,k})_{(u,j)}$}- be the { matrix} of probabilities of moves from { Stage $i-1$} for paths from group $k$. We will denote
 $$\displaystyle P^{(n)_{ik}}_{i,k}=\prod_{j=1}^{s_i}\prod_{u=1}^{s_{i-1}}p^{n^{k}_{u,i,j}}_{u,i,j,k},$$
 { which is the contribution to the likelihood of the moves}

 When a path $\ell$ from group $k$ starts, the process is checked every $b$ units of time and the moves are recorded until it terminates. It is reasonable to assume that these moves are interval censored, given the setup that we have. We don't know the exact times of the moves. We only know that exactly one move happens in an interval of length $b$. Therefore, the contribution to the likelihood of a move from Stage $i-1$ to  state $j$ of Stage $i$ within our assumptions is found by computing the probability that there is a move from $i-1$  { within time $b$ of the last record} and there is no move from Stage $i$ thereafter before $b$ time units.

If we let  $Z_1$ be the  time till move from stage $i-1$ and $Z_2$ be the time till move from stage $i$, then the contribution is

\begin{equation}
P(Z_1\leq b, Z_2>b-Z_1)=\int_{0}^{b}\int_{b-u}^{\infty}\lambda^k_i\lambda^{k}_{i+1}e^{-\lambda^k_{i}u}e^{-\lambda^k_{i+1}t}dtdu=e^{-b\lambda^k_{i+1}}\int_{0}^{b}\lambda^k_i e^{-(\lambda^k_i-\lambda^k_{i-1})u}du=\frac{\lambda^k_{i}(e^{-b\lambda^k_{i+1}}-e^{-b\lambda^k_{i}})}{\lambda^k_{i}-\lambda^k_{i+1}}.
\end{equation}
 Notice that if the consecutive stages have the same rates parameter for  $\lambda^k=\lambda$, then the above formula by considering the derivative gives the following:

\begin{equation}
P(Z_1\leq b, Z_2>b-Z_1)=\lambda^k e^{-b\lambda^k}. \label{probTrans}
\end{equation}

The above formula represents the contribution to the likelihood function of the observed interval for which a given path was seen in stage $i$ for the first time. Every stage that was visited by this path has such a contribution except for Stage 0 (nothing comes into stage 0) and Stage $m-1$ (nothing goes out of Stage $m-1$). The stages in which the path was stopped for staying too long also have this contribution (to the likelihood), and the contribution of censoring that we will provide later (nothing goes out). Considering all paths of group $k$ and the entire sample, it is clear that  arriving to stage $i$ and moving after the being recorded gives  the likelihood term

\begin{equation}
{  M_i(\blm)}=\prod_{k=1}^{K}\Big( \frac{\lambda^k_{i}(e^{-b\lambda^k_{i+1}}-e^{-b\lambda^k_{i}})}{\lambda^k_{i}-\lambda^k_{i+1}}\Big)^{v^k_i},
\end{equation}
 where $v^k_i$ is the number of paths from group $k$ that visit state $i$. When stages have equal mean waiting time $\lambda$, this becomes

\begin{equation}
M_i^e(\lambda) =e^{-b\sum_{k=1}^{K} \lambda^k v_i^k}\prod_{k=1}^{K}(\lambda^k)^{v_i^k}.
\end{equation}

Therefore, the contribution of all moves to the conditional likelihood for the data set is

\begin{equation}
M(\blm)=\prod_{i=1}^{m-2}\Big( \prod_{k=1}^{K}\Big( \frac{\lambda^k_{i}(e^{-b\lambda^k_{i+1}}-e^{-b\lambda^k_{i}})}{\lambda^k_{i}-\lambda^k_{i+1}}\Big)^{v^k_i}\Big),
\end{equation}

 or for equal average waiting time
\begin{equation}
M^e(\lambda) =e^{-b\sum_{k=1}^{K} \lambda^k \sum_{i=1}^{m-2}v_i^k}\prod_{k=1}^{K}(\lambda^k)^{\sum_{i=1}^{m-2}v_i^k}.
\end{equation}

Now, we consider the contribution of stays at the stages, excluding the time intervals { in which the  the moves took place}.
 For the state $j$ of stage $i$ and for every path $\ell$ from group $k$, we define the stay time as
  $$\displaystyle T_{i,j,k,\ell}=w^k_{i,j,\ell}b,$$
  where $w^k_{i,j,\ell}$ is the number of time intervals the given path was seen at the position after the first record there. { We will refer to this by the phrase that the path survives the move for the time $T_{i,j,k,\ell}$, which  is the time the bird has been seen in that state}.
    Let $w^k_{i,j}$ represent the number of intervals { all}  the  paths from group $k$ stayed in State $j$ of Stage $i$, and
    \begin{equation}
  w^k_{i}=\sum_{j=1}^{s_i}w^{k}_{i,j},
  \label{wki}
  \end{equation}
  be the number of intervals { all} paths from group $k$ that stayed in stage $i$.
   % We will say that the path survives the move for the time $T_{i,j,k,\ell}$.
    Thus, the conditional survival function is
\begin{equation}
\displaystyle S(T_{i,j,k,\ell})= 1-F(T_{i,j,k,\ell})=e^{-\lambda^{k}_{i+1}T_{i,j,k,\ell}}=e^{-b\lambda^{k}_{i+1}w^{k}_{i,j,\ell}}. \label{surv}
\end{equation}

Given that for the { path $\ell$ from group $k$ be seen at State $j$ of stage $i$  for }   the time $T_{i,j,k,\ell}=w^{k}_{i,j,\ell}b$, the contribution of this portion of the path to the conditional likelihood function is the survival function given by formula (\ref{surv}). Considering { all the paths in the $K$ groups  the  contribution of the stays to the likelihood is:}

\begin{equation}
S(\blm)= \prod_{i=0}^{m-2}\prod_{j=1}^{s_i}\prod_{k=1}^{K}e^{-b\lambda^{k}_{i+1}w^{k}_{i,j}}=\prod_{k=1}^{K}\prod_{i=0}^{m-2}e^{-b\lambda^{k}_{i+1}w^{k}_{i}}, \label{survt}
\end{equation}
or with equal average waiting times , it is
\begin{equation}
S^e(\lambda)=e^{-b\sum_{k=1}^{K}\lambda^{k}\sum_{i=0}^{m-2}w^{k}_{i}}, \label{survtsame}
\end{equation}

Now, we consider { the contribution to the likelihood for the paths that reach the end of the study, i.e., Stage $m-1$.} This move is different from all others because no move is expected henceforth. So its contribution will simply be the probability of moving before $b$ units of time, as given below:

\begin{equation}
L_{m-1}(\blm)=\prod_{k=1}^K \prod_{u=1}^{s_{m-1}}(1-e^{-b\lambda^k_{m-1}})^{v^k_{m-1,u}}=\prod_{k=1}^K (1-e^{-b\lambda^k_{m-1}})^{v^k_{m-1}},
\end{equation}
where $v^{k}_{m-1,u}$ is the number of visits from group $k$ to state $u$ of stage $m-1$ and $v^k_{m-1}=\sum_{u=1}^{s_{m-1}}v^{k}_{m-1,u}$.

It is clear that if we let {  $Z_{ki}$ denote the time spent at a given state of Stage $i-1$ by a path from Group $k$ }, then a censored observation  at this point has the likelihood
\begin{equation}
P(Z_{ki} > nb)=1-P(Z_{ki}\leq nb)=1-\sum_{u=1}^{s_{i}} p_{j,i,u,k}P(Z_{ki}\leq nb| \mbox{move to}\quad u)=1-\sum_{u=1}^{s_i}p_{j,i,u,k}(1-e^{-nb\lambda^k_i})=e^{-nb\lambda^k_i}.
\end{equation}
%{\cb (notion is not ideal, dependence on $k$ or $i$, may be use $Z_{ki} $ not shown on the left side of eqn. why censored obs must stay in one stage for he full time period nb??, obvious? --- it is the definition of censoring: If seen n times, then remove.)}

This shows that the effect of censoring at  any stage is the same as the effect of staying $n$ times before moving, if there was no censoring. Thus, there is no reason to consider separately the censored observations. Therefore, the conditional likelihood function is obtained as $S\times(\blm) M(\blm ) \times L_{m-1}(\blm)$ in the following formula (the conditioning is on the observed moves, denoted by $y$):

\begin{equation}
L(\blm|y)=\prod_{k=1}^{K}\prod_{i=0}^{m-2}e^{-b\lambda^{k}_{i+1}w^{k}_{i}}\prod_{i=1}^{m-2}\Big( \prod_{k=1}^{K}\Big( \frac{\lambda^k_{i}(e^{-b\lambda^k_{i+1}}-e^{-b\lambda^k_{i}})}{\lambda^k_{i}-\lambda^k_{i+1}}\Big)^{v^k_i}\Big) \prod_{k=1}^K (1-e^{-b\lambda^k_{m-1}})^{v^k_{m-1}},
\end{equation}

%{\crd need to use notation, like $Y$ for moves --- I think using "moves" is better perceived by the reader}

or for equal average waiting times
\begin{equation}
L^e(\lambda|y)=(e^{-b\sum_{k=1}^{K}\lambda^{k}\sum_{i=0}^{m-2}w^{k}_{i}})(e^{-b\sum_{k=1}^{K} \lambda^k \sum_{i=1}^{m-2}v_i^k}\prod_{k=1}^{K}(\lambda^k)^{\sum_{i=1}^{m-2}v_i^k})\prod_{k=1}^K (1-e^{-b\lambda^k})^{v^k_{m-1}}.
\end{equation}

The above derivations are obtained by multiplying over all paths of group $k$ using independence of paths. To obtain the Likelihood function for each group $k$, we need to multiply over all possible states and stages, taking into consideration the lack of memory property of the exponential distribution. We need now to multiply by the transition probabilities of the states to obtain the full likelihood of the data:

\begin{equation}
L(\bold \lambda|y)=\prod_{k=1}^{K}\prod_{i=0}^{m-2}e^{-b\lambda^{k}_{i+1}w^{k}_{i}}\prod_{i=1}^{m-2}\Big( \prod_{k=1}^{K}\Big( \frac{\lambda^k_{i}(e^{-b\lambda^k_{i+1}}-e^{-b\lambda^k_{i}})}{\lambda^k_{i}-\lambda^k_{i+1}}\Big)^{v^k_i}\Big) \prod_{k=1}^K (1-e^{-b\lambda^k_{m-1}})^{v^k_{m-1}} \prod_{i=1}^{m-1}P_{i,k}^{(n)_{ik} },
\end{equation}

giving for the case of equal average waiting times
\begin{equation}
L^e(\lambda|y)=(e^{-b\sum_{k=1}^{K}\lambda^{k}\sum_{i=0}^{m-2}w^{k}_{i}})(e^{-b\sum_{k=1}^{K} \lambda^k \sum_{i=1}^{m-2}v_i^k}\prod_{k=1}^{K}(\lambda^k)^{\sum_{i=1}^{m-2}v_i^k})\prod_{k=1}^K (1-e^{-b\lambda^k})^{v^k_{m-1}}P_{ik}^{(n)_{ik} }.  \label{mle}
\end{equation}

\section{Maximum likelihood estimation }

Considering the above likelihood function, it is clear  that the maximum likelihood estimator { (MLE)} of the parameters exists and is unique. For the parameter $\lambda$, there is no closed form for the MLE but it can be estimated numerically provided that the data is given. Using $n_{u,i}^{ k}=\sum_{j=1}^{s_{i}}n_{u,i,j}^{ k}$ as number of moves from state $u$ of stage $i$ for paths from group $k$, the transition probabilities have estimators

\begin{equation} \label{MLEK}
 \hat{p}_{u,i,j}^{k}=\frac{n^{k}_{u,i,j}}{n_{u,i}^{k}}.
\end{equation}

%%%%%%%%%%%%%%

\subsection{Distributional technicalities}

{ In this section, we consider the probability distributions of some quantities of interest, such as the number of times a path is seen at a given location. We shall find the probability to reach a given state or find a match.}

Notice that the distribution of $T^k_{i,\ell}$ (time spent by the $\ell^{th}$ path of group $k$ in Stage $i$) is not exponential. It is a more complex distribution because the move is not recorded directly after time $T=nb$.  Given that we are already at stage $i-1$, this time is $T^k_{i,\ell}=Z_1+bw^{k}_{i,j,\ell}+(1-\delta_\ell)Z_2$, where $Z_1,$ $Z_2$ have  independent exponential distributions { with common parameter $\lambda^{k}$}, { $w^{k}_{i,j,\ell}$ is a discrete random variable representing the number of { time intervals}  the path was seen at this { stage} with }  values between 0 and n, and $\delta_\ell$ is the indicator of removal of the individual { $\ell$}  from the study { in Stage $i$} (equals 1 when no move is made to the next stage, and 0 otherwise).
%This distribution will use {\cb all possible values of   $w^{k}_{i,j,\ell}$}, $\{0,1, 2, \cdots, n\}$.
 Let $N_{i,j,\ell}^k$ be the number of records at { State $j$ of Stage $i$ for  path $\ell$ from group $k$}, { and thus $N_{i,j,\ell}^k=  w^{k}_{i,j,\ell}+1$} given that the path has reached state $j$ of Stage $i$.  { Note that the conditional distributions of these random variables do not depend on $j$, since we have the same average waiting time across stages.} It is clear that { $N_{i,j,\ell}^{k}=\gamma_{ij} (1+w^k_{i,j,\ell})$, where $\gamma_{ij}$ } is the indicator of the path reaching state $j$ of Stage $i$ ($\gamma_{ij}=1$ if the path reaches state $j$ of State $i$, equals $0$ otherwise). Thus,  the conditional distribution of this random variable is obtained in the following way (conditional on the path reaching Stage $i$).

\begin{equation}
P(w^{k}_{i,j,\ell}=v|  \gamma_{ij} =1 )=e^{-vb\lambda^{k}}(1-e^{-b\lambda^{k}})(1-\delta_\ell)I_{\{0,...,n\}}(v) +\delta_\ell e^{-nb\lambda^{k}}. \label{probmove}
\end{equation}
%{\crd[[ the above eqn may be written to read better, looks like in one case $v<n$..... Yes, but I didn't want another formula with cases]], also need to use $\lambda^{(k)}$ or $\lambda_k$ - this keeps consistency with $\lambda_i^{k}$}
%{(\cb I have modified the formula. When $v=n$, the path is automatically removed from the study. Equivalent to $\delta_{\ell}=1$)}

{ Since the distribution of the quantities $w^{k}_{i,j,\ell}$and $N_{i,j,\ell}^k$ do not depend on the specific $j$ or $\ell$, we will henceforth simplify their notation and write $w^{k}_{i}$ and  $N_{i}^k$. }

 Let $P_k (i'+1|i')$ be the probability that a given path reaches  stage $i'+1$
 and not move within the last $b$ units of time given that the path is seen for the first time at Stage $i'$. %{\cb [[ I think we are on the same page - so I have modified]]}
%  This probability is equal to   the probability of spending less than $n$ units of time at Stage $i'$, which is  (from  formula (\ref{probmove}) times . {\crd [[ this last sentence is right?? It is ok. ]]} {\crd [[isn't the probability of spending less than $n$ units of time at Stage $i'$ is $(1-e^{-bn\lambda^k})$? It is. That's what you get by adding all $v$ less than $n$ of the given formula]]}
 Since the probability of spending { less than $n$ segments of $b$ units of time at Stage $i'$ is $1-e^{-bn\lambda^k}$  (from  formula (\ref{probmove}) ),} and the probability of getting to Stage $i'+1$ and not moving within $b$ units of time is $\lambda^k e^{-b\lambda^k}$, by formula (\ref{probTrans}), we obtain

\begin{equation}
P_k (i'+1|i')=\lambda^k (1-e^{-bn\lambda^k})e^{-b\lambda^k}.
\end{equation}

Thus, if we denote $P_k (i)$ the probability that a given path from group $k$ is seen at stage $i$ one time, we have for $i<m-1$
\begin{equation}
P_k(i)=\prod_{i'=0}^{i-1}P_k (i'+1|i')=(\lambda^ke^{-b\lambda^k}(1-e^{-nb\lambda^k}))^{i}.
\end{equation}

{ If we denote { $N^{k}_{m-1}$}- the Bernouili  variable that indicates the path $\ell$ reaching stage $m-1$, then the probability that the path reaches stage $m-1$ (i.e., the female picks a male) is }
%{\crd  [[ Is $N^{k}_{m-1}$ defined somewhere?, =1 if path $\ell$ reaches m-1? This is it defined here. The Stage m-1 is a match. It is just singled out for its nature.]]}

\begin{equation}
P(N^{k}_{m-1}=1)=P_k(m-1)=(\lambda^ke^{-b\lambda^k}(1-e^{-nb\lambda^k}))^{m-2}(1-e^{bn\lambda^k}).
\label{pkm}
\end{equation}
  Note that $ P_k(m-1)$ is not the same as $P_k(i)$ at $i=m-1$, and that the latter is defined only for $i<m-1$.
  % {[[ \crd not clear to me why it needs to b seen at stage m-2 ---- m-1 is the last Stage to which they get from not being censored in Stage $m-2$]] }
Reaching the final stage is different from other stages because we don't need to consider time after reaching the final stage. The process stops  when the the male is chosen. So, the obtained probability is the product of the probabilities to reach $m-2$ and not move within $b$ units of time and the probability of reaching the final stage after being seen on the previous stage.

{ Artificially setting $w^{k}_{m-1}=-1$  (see (\ref{wki})}, for the case when the path doesn't reach stage $i$,
the probability that the path will be stopped prior to reaching stage $m-1$ (meaning not get a match in male choices) is
\begin{equation}
P(w^{k}_{m-1}=-1)=1-(\lambda^ke^{-b\lambda^k}(1-e^{-nb\lambda^k}))^{m-2}(1-e^{bn\lambda^k}).
\end{equation}

Combining the above equations for $i<m-1$, we obtain,

\begin{equation}\label{Nik}\displaystyle
P(w_{i}^{k}= v)=\begin{cases} e^{-vb\lambda^{k}}(1-e^{-b\lambda^{k}})(\lambda^ke^{-b\lambda^k}(1-e^{-nb\lambda^k}))^{i},\quad \mbox{for}\quad v=0,1,\cdots n-1, \\
e^{-nb\lambda^{k}}(\lambda^ke^{-b\lambda^k}(1-e^{-nb\lambda^k}))^{i}, \quad \mbox{for}\quad v=n,\\
1-(\lambda^ke^{-b\lambda^k}(1-e^{-nb\lambda^k}))^{i} \quad \mbox{for} \quad v=-1\\
0, \quad \mbox{otherwise}.
\end{cases}
\end{equation}

\subsection{Mean and variance of { $N_{i}^{k}$} }

%{\crd $N_{i,\ell}^{k}$ defined somewhere, \# of times path $\ell$ seen at Stage $i$? defined above formula 18}

%{\crd [[ It looks like $N_{i}^{k}$ and $w_{i}^{k}$ are the same, based on what is meant as "records" in an earlier section, it is the number of times path is seen. Can you clarify the difference somewhere appropriate? ... There is a difference of 1 between the two.]]}

\subsubsection{ Conditional mean and variance of $N_{i}^{k}$}
Given that { a path} has reached stage $i$, { we need to find the conditional mean and variance ( $\mathbb{E}_i(N_{i}^k)$ and $var_{i}(N^k_{i})$ respectively) of  the total number of records at stage $i$, for $i< m-1$. }
%We denote $\mathbb{E}_i(N_{i}^k)$ and $var_{i}(N^k_{i})$ these conditional mean and variance for $i<m-1$.
Using the distribution obtained in formula (\ref{Nik}) and the fact that on the condition that the path is in Stage $i$, $N_i^k=w_{i}^k$, %{\crd [[this formula is unconditional dist]] --- for $w$, but $(N|\gamma_i=1) = 1+w$},
we get,
%{\cb I meant that the formula does not directly apply and better to give a better explanation, ....but agree with the formula below]]}
\begin{align*}
\mathbb{E}_ i(N^k_{i}) = & \sum_{u=0}^{n-1}(u+1)e^{-ub\lambda^{k}}(1-e^{-b\lambda^{k}})+(n+1)e^{-nb\lambda^{k}}\\
{=} & \sum_{u=0}^{n-1}ue^{-ub\lambda^{k}}(1-e^{-b\lambda^{k}})+\sum_{u=0}^{n-1}e^{-ub\lambda^{k}}(1-e^{-b\lambda^{k}})+(n+1)e^{-nb\lambda^{k}}\\
{=} & \sum_{u=0}^{n-1}ue^{-ub\lambda^{k}}(1-e^{-b\lambda^{k}})+1+ne^{-nb\lambda^{k}}.
\end{align*}

Notice that, if we set $\displaystyle h(\lambda^{k})=\sum_{u=0}^{n-1}e^{-ub\lambda^{k}}=\frac{1-e^{-bn\lambda^{k}}}{1-e^{-b\lambda^{k}}}$, then

\begin{equation*}
\sum_{u=0}^{n-1}ue^{-ub\lambda^{k}}(1-e^{-b\lambda^{k}})=\frac{-1}{b}h'(\lambda^{k})(1-e^{-b\lambda^{k}})=-\frac{ ({1-e^{-b\lambda^{k}}})}{b}\frac{\partial}{\partial \lambda^{k}}\large(\frac{1-e^{-bn\lambda^{k}}}{1-e^{-b\lambda^{k}}}\large).
\end{equation*}

Therefore, we obtain $\displaystyle \mathbb{E}_ i(N^k_{i })=-\frac{ ({1-e^{b\lambda^{k}}})}{b}\frac{\partial}{\partial \lambda^{k}}\Large(\frac{1-e^{-bn\lambda^{k}}}{1-e^{-b\lambda^{k}}}\Large)+1+ne^{-nb\lambda^{k}}.$
%{\crd [[Did a slight reorganization in the formulas. Also removed suffix $\ell$ in the notation for  $N$ up to the point this round of mine goes]] ok}
Working out the last equality leads to
\begin{equation}
\mathbb{E}_ i(N^k_{i })=\frac{1-e^{-b(n+1)\lambda^{k}}}{1-e^{-b\lambda^{k}}}.
\end{equation}
To obtain the conditional variance, we need to compute $\mathbb{E}_i ((N_{i }^k)^2 )$. Using the same arguments that we have applied for $\mathbb{E}_i (N_{i }^k)$, we get
\begin{align*}
\mathbb{E}_ i ((N^k_{i })^2) =& \sum_{u=0}^{n-1}(u+1)^2e^{-ub\lambda^{k}}(1-e^{-b\lambda^{k}})+(n+1)^2e^{-nb\lambda^{k}}\\
{=} & \sum_{u=0}^{n-1}(u+1)e^{-ub\lambda^{k}}(1-e^{-b\lambda^{k}})+(n+1)e^{-nb\lambda^{k}}+\\ {} & +\sum_{u=0}^{n-1}u(u+1)e^{-ub\lambda^{k}}(1-e^{-b\lambda^{k}})+n(n+1)e^{-nb\lambda^{k}}.
\end{align*}

It turns out that
\begin{equation}
\mathbb{E}_ i ((N^k_{i })^2) =\mathbb{E}_ i(N^k_{i })-\frac{(1-e^{-b\lambda^{k}})}{b}\frac{\partial}{\partial\lambda^{k}}(\frac{\mathbb{E}_ i(N^k_{i })-(n+1)e^{-nb\lambda^{k}}}{1-e^{-b\lambda^{k}}})+n(n+1)e^{-nb\lambda^{k}}.
\end{equation}
Simple calculations lead to
\begin{equation*}
\mathbb{E}_ i ((N^k_{i })^2)=\frac{1+e^{-b\lambda^{k}}-(2n+3)e^{-(n+1)b\lambda^{k}}+(2n+1)e^{-(n+2)b\lambda^{k}}}{(1-e^{-b\lambda^{k}})^2}.
\end{equation*}

Therefore, the conditional variance is
\begin{equation} var_i (N_{i }^k)= \frac{e^{-b\lambda^{k}}(1-(2n+1)(e^{-nb\lambda^{k}}-e^{-(n+1)b\lambda^{k}})-e^{-(2n+1)b\lambda^{k}})}{(1-e^{-b\lambda^{k}})^2}.
\end{equation}
One can check { using L'Hopital's rule}
%derivatives with respect to $\lambda^{k}$ at $\lambda^{k}=0$
that for $\lambda^{k}=0$, $\mathbb{E}_ i ((N^k_{i })^2)=(n+1)^{2}$, $\mathbb{E}_ i (N^k_{i })=n+1$ and $var_{i}(N_{i }^k)=0$. This is the case when there is actually no possible move once the path gets to $i$. The path in this case will certainly be recorded $n+1$ times at stage $i$ and the process will stop. On the other hand, as $n\to\infty$, $N_{i }^k$ converges in distribution to the geometric distribution with probability of success $p_i^k=1-e^{-b\lambda^{k}}$.

\subsubsection{Mean and variance of $N_{i }^k$}

Computing the  mean by conditioning first on reaching stage $i$ and not moving within the first $b$ units of time, we obtain
\begin{eqnarray}\displaystyle
\mathbb{E} (N^k_{i })=\mathbb{E}_ i (N^k_{i })\cdot P_k(i)=\frac{1-e^{-b(n+1)\lambda^{k}}}{1-e^{-b\lambda^{k}}}\cdot (\lambda^ke^{-b\lambda^k}(1-e^{-nb\lambda^k}))^{i},\\
\mathbb{E} ((N^k_{i })^2)=\frac{1+e^{-b\lambda^{k}}-(2n+3)e^{-(n+1)b\lambda^{k}}+(2n+1)e^{-(n+2)b\lambda^{k}}}{(1-e^{-b\lambda^{k}})^2}\cdot (\lambda^ke^{-b\lambda^k}(1-e^{-nb\lambda^k}))^{i}.
\end{eqnarray}
The variance can therefore be obtained as $\displaystyle var(N^k_{i })=\mathbb{E}((N^k_{i })^2)-(\mathbb{E}(N^k_{i }))^2$.

These formulas can be used to make statistical inference about the number of stays at a given stage. It is important to notice that the mean depends on $b$ and $\lambda^k$. Thus, in a setup where inference on $\lambda^k$ is known, one can choose the appropriate $b$ for a given target result.

\subsection{Inference on { the probability of} reaching the final stage from a given position}

{ As in (\ref{pkm})}, the probability to reach stage $m-1$ for a  path from group $k$ is
\begin{equation}
p_k(m-1)=(\lambda^ke^{-b\lambda^k}(1-e^{-nb\lambda^k}))^{m-2}(1-e^{bn\lambda^k}).
\label{pkm-1}\end{equation}
Similarly, it is derived that the probability to reach the final stage from any given stage $i$ is
$$p_{ki}(m-1)=(\lambda^ke^{-b\lambda^k}(1-e^{-nb\lambda^k}))^{m-2-i}(1-e^{bn\lambda^k}).$$

 Recall that the MLE of $\lambda^k$ was discussed earlier using formula (\ref{mle}). The regularity conditions being satisfied, it is clear that the CLT holds for $\hat{\lambda}^k$ (MLE of $\lambda^k$) in the form

\begin{equation}
{ \sqrt{ N_k}(\hat{\lambda}^k-\lambda^k) \to N(0, \frac{1}{I(\lambda^k)}),}
\end{equation}
where $I(\lambda^k)$ is the Fisher information of $\lambda^k$. Taking into account the fact that $p_k$ { (either  $p_k(m-1)$ or $p_{ki}(m-1) $) is  a function of $\lambda_k$}, it can be concluded using the Delta method that
\begin{equation}
{ \sqrt{ N_k}(p_k(\hat{\lambda}^k)-p_k) \to N(0, \frac{(p_k'(\lambda^k))^2}{I(\lambda^k)}),}
\end{equation}
where $p_k(\hat{\lambda}^k)$ is the MLE of $p_k$ that is obtained by plugging the MLE of $\lambda^k$ into the formula of $p_k$ and $p_k'$ is the derivative of $p_k'$ with respect to $\lambda^k$. { Similar result holds for the common $p_k$ for the case of equal $\lambda$ across $K$ groups. These results can be used to obtain approximate confidence intervals for the probability $p_k$'s.}

For $N_k$ observations from group $k$,  { let $N_k(m)$ be the number of paths that reach the final stage.  The random variable $N_k(m)$ is the sum of $N_k$ independent and identically distributed Bernoulli random variables, and has a binomial distribution with parameters $\displaystyle (N_k, p_k(m-1))$. }
%(\lambda^ke^{-b\lambda^k}(1-e^{-nb\lambda^k}))^{m-2}(1-e^{bn\lambda^k})). $
%The random variable $N_k(m)$ is the sum of $N_k$ independent and identically distributed Bernoulli random variables.
From here on, we will use $p_{k}$ and $p_{ki}$ for $p_{k}(m-1)$ and $p_{ki}(m-1)$ when there is no possible ambiguity.  From the above fact, it is clear that the total number of paths that reach the final stage $\displaystyle N(m)=\sum_{k=1}^{K}N_{k}(m)$ is the sum of independent binomial random variables. As $N_k\to \infty$, the central limit theorem (CLT) holds in the form

\begin{equation}
(\sqrt{N_k}\frac{(\frac{ N_{k}(m)}{N_k}-p_k)}{\sqrt{p_k(1-p_k)}}, k=1,\cdots K)\to N(\underline{0}_K, I_K),
\end{equation}
where $\underline{0}_K$ is the zero vector of size $k$, $I_K$ is the $K\times K$ identity matrix and $N(\underline{0}_K, I_{K})$ is the standard K-dimensional normal distribution.

This general form of the CLT can not be used to find the limiting distribution of $N(m)$. If we consider a balanced design, in which the same number of paths is selected from each of the groups ($N_k=N$), then it follows that
\begin{equation}
\sqrt{N} (\frac{N(m)}{N}-\sum_{k=1}^{K}p_k) \to N(0,  \sum_{k=1}^{K}p_k(1-p_k)). \label{cltNm}
\end{equation}
\indent The central limit theorem (\ref{cltNm}) can be used to obtain inference, { such as prediction intervals,} for  the total number of paths that reach the final stage, using the MLE of the parameters that are involved.
To obtain results for reaching the final stage from a given position $i$, it is enough to replace  $p_k$ by $p_{ki}$ in the formula and use the appropriate MLE.

\subsection{Inference on reaching a particular final state $u$}
It is clear that the probability of reaching the final state $u$ is a product of the probability of reaching the final stage and the probability of reaching the specific state $u$ from stage $m-2$. The latest is the sum of the probabilities of reaching this state from each of the states of stage $m-2$. Thus, if we denote $P_{ki}(u)$ the probability of reaching final state $u$ from a given state $i$, we obtain

\begin{equation}
p_{ki}(u)=\sum_{j=1}^{s_{m-2}}p^k_{j,m-1, u}(\lambda^ke^{-b\lambda^k}(1-e^{-nb\lambda^k}))^{m-2-i}(1-e^{bn\lambda^k}).
\end{equation}

Maximum likelihood estimators of these probabilities are obtained by using the estimates from the above MLE of model parameters. These probabilities can be used to define test statistics for differences between the groups $k$ on one hand and to have inference on preferences or classification of the final states (males) for each of the groups (females). Using this method of comparison, we can conclude that the ranking of states $ u= (1,\cdots, s_{m-1})$ (males) for group $k$ is that of the quantities

\begin{equation}
 K_{uk} =\sum_{j=1}^{s_{m-2}}p^k_{j,m-1, u}, \quad u=1, \cdots, s_{m-1}.
\end{equation}

Recall that the likelihood estimators of $(p_{j,m-1, u}^k, 1\leq u\leq s_{m-1}, 1\leq j\leq s_{m-2})$ are $\hat{p}_{j, m-1, u}^k=\frac{n_{j, m-2, u}^{k}}{n^{k}_{m-2, j}}$. Notice that these { estimates are based on  counts  from multinomial distributions} conditioned on the number of observations that have reached the states $j$. Thus, the conditional CLT holds for these variables in the form
\begin{equation}
\frac{(n_{j, m-2, 2}^{k},\cdots, n_{j,m-2, s_{m-1}})}{\sqrt{n^k_{m-2,j}}} \longrightarrow {\bf N} ({\bf P}^k, {\bf \Sigma}^k) , \quad {\bf P}^k=(p^k_{j,m-1,2}, \cdots, p^{k}_{j,m-1,  s_{m-1}}), \quad \mbox{and}\quad {\bf \Sigma}^k= diag {\bf P}^k-{\bf P}^{kT} {\bf P}^k.
\end{equation}

\subsubsection{MLE inference for classification of males within each group of females}

Considering The MLE of $K_{uk}$, we have $$ \displaystyle \hat{K}_{uk}=\sum_{j=1}^{s_{m-2}} \frac{n^k_{j,m-2,u}}{n^{k}_{m-2,j}}.$$

Using the fact that paths from different groups are independent and the sum of independent normal distributions is a normal distribution, we obtain
\begin{equation*}
\hat{K}_{uk}-K_{uk}=\sum_{j=1}^{s_{m-2}}( \frac{n^k_{j,m-2,u}}{n^{k}_{m-2,j}}-K_{uk})\to \sum_{j=1}^{s_{m-2}}\frac{1}{\sqrt{n^k_{m-2,j}}}N(0, p^k_{j,m-1,u}(1-p^k_{j,m-1,u}))
\end{equation*}
Therefore,
$\displaystyle
\hat{K}_{uk}-K_{uk}\to N(0, \sum_{j=1}^{s_{m-2}}\frac{p^k_{j,m-1,u}(1-p^k_{j,m-1,u})}{n_{m-2,j}^k}),
$
and
$\displaystyle
(\hat{K}_{uk}-K_{uk}, \quad u=1, \cdots, s_{m-1}) \to {\bf N}(0, \sum_{j=1}^{s_{m-2}}\frac{1}{n^k_{m-2,j}}{\bf \Sigma}^k).
$

If we assume that $n^k_{(m-2),j}=n^k_{m-2}b_{j}$ with $b_1+\cdots+b_{s_{m-2}}=1$ (this is a plausible assumption because the mean numbers the reach states are proportional to the number of observations in stage $m-2$), the it follows that

\begin{equation}
\sqrt{n^k_{m-2}} (\hat{K}_{uk}-K_{uk}, \quad u=1, \cdots, s_{m-1}) \to {\bf N}(0, \sum_{j=1}^{s_{m-2}}\frac{1}{b^k_{j}}{\bf \Sigma}^k).
\end{equation}

\subsubsection{Substitute inference for classification of males within each group of females}

 Considering the estimators of $K_{uk}$ as $\displaystyle \tilde{K}_{uk}=\frac{n^{k}_{m-2,u}}{n^{k}_{m-2}}$ and setting $K_k=(K_{uk}, u=1, \cdots, s_{m-1})$, arguments similar to the above lead to the central limit theorem in the form

\begin{equation}
{\sqrt{n^{k}_{m-2}}}\Big( \tilde{K}_{uk}, u=1, \cdots, s_{m-1}\Big) \longrightarrow {\bf N} \Big( K_{k}, {\bf \Sigma}^k\Big). \label{cltKtilde}
\end{equation}

Moreover, these are unbiased estimators. Therefore, conditions of Theorem 1 of Chen and Szroeter (2014) are satisfied for the parameters $\mu_j=K_{j+1}-K_{j}, j=1,\cdots, u-1$. Assume that the null hypothesis is $$H_0: \mu_j\ge 0, j=1,\cdots, u-1  \quad vs \quad H_1: \quad not \quad H_0. $$
Let $\psi_n (x)=\psi(K(n)x)$, $\Psi=(\psi_n (\theta_j\mu_j),\quad j=1, \cdots u-1)^{T}$, $\Lambda_{n}(\mu_j , v_{jj})=v_{jj}\psi_n (\mu_j )K(n)n^{-1/2}$ and $\Lambda=(\Lambda_n (\theta_j \mu_j, \theta_j^2v_{jj}), j=1,\cdots, u-1 )^{T}$. Define $Q_1 =\sqrt{n}\hat{\Psi}^{T}\hat{\Delta}\hat{\mu}-(1,\cdots, 1)\cdot \hat{\Lambda}$, $Q_2=\sqrt{\hat{\Psi}^{T}\hat{\Delta}\hat{V}\hat{\Delta}\hat{\Psi}}$, where hats substitute estimators for the variables and $(\theta_i, i=u-1)$ is the vector of inverse standard deviations of $(\hat{\mu}_i, i=1, \cdots u-1)$. Notice that these vectors can be formed after finding estimates of the $K's$ to have an informed guess of the inequalities that are needed in the hypotheses.

Using these notations, the test statistic is defined as

$$
Q=\begin{cases} \Phi(\frac{Q_1}{Q_2}), \quad \mbox{if}\quad Q_2>0,\\ 1, \quad \mbox{if} \quad Q_2=0.
\end{cases}
$$

{ Now, we provide a list of conditions for use in the following theorem about the test statistic $Q$. }
\begin{align}
V &= var(\hat{\mu}), \quad \hat{V}\rightarrow^P V  \label{cond1}\\
\Delta &= diag(\theta_1, \cdots, \theta_{u-1}), \quad   \hat{\Delta}\rightarrow^P \Delta
\end{align}
\begin{align}
\psi(x)&=1-F(x),\quad F(x) \quad \mbox{is a CDF with continuous and bounded pdf with} \quad F(0)\ne 1\\
K(n) &\quad \mbox{is increasing, positive, and} \quad \lim_{n\to \infty}K(n)=\infty, \lim_{n\to\infty}n^{-1/2}K(n)=0\\
&\lim_{n\to \infty}\sqrt{n}\psi(K(n)x)=0 \quad \mbox{for } \quad x>0 \label{condend}
\end{align}

Chen and Szroeter (2014) conclude the following for our problem.
\begin{theorem} \label{Theo1}
For our example, under $H_0$, if (\ref{cond1})-(\ref{condend}) are satisfied and the estimator of $\mu$ satisfies the CLT with positive definite $V$, then $Q\rightarrow^P 1$.
\end{theorem}

For a modified null hypothesis $\displaystyle H_0^{\prime} : \mu_j\ge 0, j=1,\cdots t-1, t+1,\cdots, u-1, \mu_t=0  \quad vs \quad H_1: \quad not \quad H_0. $

\begin{theorem}\label{Theo2}
Under $H_0^{\prime}$,  if (\ref{cond1})-(\ref{condend}) are satisfied and the estimator of $\mu$ satisfies the CLT with positive definite $V$, then $Q\longrightarrow U(0,1)$ and $H_0^{\prime}$ would be rejected at the level $\alpha$ if $Q<\alpha$.
\end{theorem}

If Theorem \ref{Theo1} can be used to determine what differences to consider for the test, Theorem \ref{Theo2} can be used to determine which choices are similar. Failling to reject $H_0^{\prime}$ would indicate that $\mu_t$ can't be statistically different from 0. Thus, the two groups that are related to it would be similar.

\subsection{The most probable path as differentiation tool for groups of females}

Considering the most probable paths.
For each observation from group $k$, there are $u\times m$ paths. Each of the paths has its own probability that can be referred to as preference. We can consider the probability of the preferred path of the group as an indicator of choice. In this case, each of the groups can be identified by its most probable paths. Two groups will be deemed similar if they share the same most probable path.

Based on the model with waiting times independent of the state the move gets into, it is easy to show that the most probable path will consist of most probable possible consecutive moves. Thus, for an observation from group $k$, the most probable path has probability
 $$P(mk)= \prod_{i=1}^{m-1} \max(p^k_{j, i-1,u}, 1\leq u\leq s_{i}, 1\leq j\leq s_{i-1})(1-e^{-bn\lambda^{k}})^m .$$

We can have a maximum likelihood for this probability and perform a test for differences of groups, using the above methodology. In doing this, we need to carefully sequentially test each of the statements to find the maximum at each of the states for each of the groups, before comparing the most probable paths for groups.

\section{Discussion}
This work presents a framework of the probability theory useful for a rigorous statistical approach to study the problem of male choices important to researchers in the area, that is also applicable to similar situations, where the  scientist can not be present to record the events, but has fixed windows to check the system. The model assumes that not more than one event is possible in each of the windows. The assumption is driven by practical needs of the system to generate the moves.

The model is rather complex making closed form maximum likelihood estimation  untenable, because the observer doesn't see the time of the moves. In a system where the actual time of the move can be seen, the complexity of the problem and hence the model would be reduced. We will propose this simplified model as text-book example in a subsequent developments of this theory.

This work will lay the background for applications in biology, econometrics, queuing theory and other fields. While the biological motivation is provided here, in econometrics, the model can be used for classification of means when the exponential model is assumed and in queuing theory many options for applications are available, including but not limited to providing several services to each of the clients in an office with a supervisor calling in from time to time the clients when too much time is spent on him.

\section*{Acknowledgements} Dr. Longla has been supported by the CLA of the University of Mississippi Summer Grant for this work.

\end {document}